\documentclass[12pt]{article}

\usepackage{amssymb}
\usepackage{hyperref}

\begin{document}

\newtheorem{0}{DEFINITION}[section]
\newtheorem{1}{LEMMA}[section]
\newtheorem{2}{THEOREM}[section]
\newtheorem{3}{COROLLARY}[section]
\newtheorem{4}{PROPOSITION}[section]
\newtheorem{5}{REMARK}[section]
\newtheorem{6}{EXAMPLE}[section]
\newtheorem{7}{ALGORITHM}[section]
\newtheorem{8}{CONJECTURE}[section]
\newtheorem{9}{QUESTION}[section]

\title{HOMC: A MATLAB Package for Higher Order Markov Chains}

\author{
Jianhong Xu\thanks{School of Mathematical and Statistical Sciences, Southern Illinois University Carbondale, Carbondale, IL 62901, USA. Email: \texttt{jhxu@siu.edu}} 
}

\maketitle

\begin{abstract}
We present a MATLAB package, which is the first of its kind, for Higher Order Markov Chains (HOMC). It can be used to easily compute all important quantities in our recent works relevant to higher order Markov chains, such as the $k$-step transition tensor, limiting probability distribution, ever-reaching probability tensor, and mean first passage time tensor. It can also be used to check whether a higher order chain is ergodic or regular, to construct the transition matrix of the associated reduced first order chain, and to determine whether a state is recurrent or transient. A key function in the package is an implementation of the tensor ``box'' product which has a probabilistic interpretation and is different from other tensor products in the literature. This HOMC package is useful to researchers and practitioners alike for tasks such as numerical experimentation and algorithm prototyping involving higher order Markov chains.
\end{abstract}

\noindent {\bf keywords}: higher order Markov chains, MATLAB, tensors, transition probabilities, ever-reaching probabilities, limiting probability distributions, mean first passage times, classification of states

\section{Introduction}

Let $X=\{X_t : t=1,2,\ldots\}$ be a stochastic process, where $t$ and $X_t$ can be regarded as time and the state of a system at time $t$, respectively. For convenience, the states of the system are often labeled as $1$ through $n$ with $n \ge 2$. The set $S=\{1, 2, \ldots, n\}$ is called the state space. Then, this process $X$ is called a first order Markov chain if the future state $X_{t+1}$ of the system can be determined by its present state $X_t$, meaning 
$$\Pr(X_{t+1}=i | X_t=j, \ldots, X_1=k)=\Pr(X_{t+1}=i | X_t=j)$$
for all $t \ge 1$ and $i, j, \ldots, k \in S$. In addition, the chain is called homogeneous if the above probability depends only on $i$ and $j$ but is independent of $t$. Denote $p_{ij}=\Pr(X_{t+1}=i | X_t=j)$, which is called the transition probability from states $j$ to $i$. The $n$ by $n$ matrix $P=[p_{ij}]$ is called the transition matrix of the chain. The matrix $P$ is stochastic, that is, $0 \le p_{ij} \le 1$ for all $i, j \in S$ and $\sum_{i \in S}p_{ij}=1$ for all $j \in S$.

On the other hand, a higher order Markov chain arises if the future state of the system depends on not only its present state but also one or more of its past states. Specifically, on letting $m \ge 3$, the process $X$ is called a higher order Markov chain of order $m-1$, or simply an $(m-1)$th order Markov chain, if 
\begin{equation}
\label{mark}
\begin{array}{c}
\Pr(X_{t+1}=i_1 | X_t=i_2, \ldots, X_{t-m+2}=i_m, \ldots, X_1=i_{t+1})\\
=\Pr(X_{t+1}=i_1 | X_t=i_2, \ldots, X_{t-m+2}=i_m)
\end{array}
\end{equation}
for all $t \ge m-1$ and $i_1, i_2, \ldots, i_m, \ldots, i_{t+1} \in S$. In particular, such a higher order chain reduces to the first order case if $m=2$. Similar to the first order one, the higher order chain is called homogeneous if the probability in (\ref{mark}) depends only on $i_1, i_2, \ldots, i_m$ yet is independent of $t$. Denote $p_{i_1i_2\ldots i_m}=\Pr(X_{t+1}=i_1 | X_t=i_2, \ldots, X_{t-m+2}=i_m)$, which is called the transition probability from states $(i_2, \ldots, i_m)$ to $i_1$. The $m$th order, $n$ dimensional tensor $\mathcal{P}=[p_{i_1i_2\ldots i_m}]$ is called the transition tensor of the higher order chain. This tensor $\mathcal{P}$ is stochastic too in the sense that $0 \le p_{i_1i_2\ldots i_m} \le 1$ for all $i_1, i_2, \ldots, i_m \in S$ and $\sum_{i_1 \in S}p_{i_1i_2\ldots i_m}=1$ for all $i_2, \ldots, i_m \in S$. Throughout this article, a Markov chain of any order is always assumed to be homogeneous and is often referred to simply as a chain.

Higher order chains have been the focus of many recent works, see \cite{Chang13, Geiger17, Han22, Han24a, Han24b, Han25, Hu14, Li16, Li14, Wu17} and the references therein. They have also appeared in a variety of applications \cite{Baena18, Deneshkumar19, Flett16, Gleich15, Ky18, Islam06, Lan13, Liu20, Masseran15, Sanjari17, Xiong19, Yang20}. To the best of our knowledge, however, there is no dedicated MATLAB package for higher order chains. The goal of this article is to present for the first time such a package, named by the acronym HOMC for Higher Order Markov Chains.

MATLAB is a high-level computing environment that has been widely adopted by researchers and practitioners in various fields as a powerful tool for numerical experiment, simulation, and visualization. When it comes to higher order chains, MATLAB is particularly attractive because of its capability of dealing with multidimensional data such as the transition tensor $\mathcal{P}$. To store $\mathcal{P}$ in MATLAB, for example, we can easily achieve that by assigning the respective transition probabilities to each frontal slice \cite{Martin13} $\mathcal{P}(:,:,i_3, \ldots, i_m)$, where $i_3, \ldots, i_m \in S$. These frontal slices are $n$ by $n$ matrices and hence can be conveniently handled and manipulated in MATLAB.

In our HOMC package, a key function is an implementation of the $\boxtimes$, or ``box'', product of $m$th order, $n$ dimensional tensors \cite{Han24a, Han25}. This operation has a useful probabilistic interpretation and is different from other types of existing tensor multiplications \cite{Kolda09, Qi17}. The $\boxtimes$ product may be directly implemented by first constructing all $(i_1, i_2, \ldots, i_m)$ for $i_1, i_2, \ldots, i_m \in S$. This approach clearly requires a huge amount of storage memory and, therefore, becomes impractical when the size of the chain is large. Accordingly, we have taken a different approach by exploiting MATLAB's built-in function {\tt ind2sub} to generate $(i_1, i_2, \ldots, i_m)$ on the fly. The same idea has been used across the other functions in the package when necessary so as to cut on their memory usage. Overall, the main storage utilization of the package is dominated by that of the transition tensor $\mathcal{P}$ and, if applicable, output tensors of the same size. Thus, its space complexity is $O(n^m)$. This is the best that can be accomplished in coping with a dense tensor problem such as general higher order chains.

As seen in their definitions, a higher order chain reduces to a first order one when $m=2$. Consistent with that, our HOMC package can also deal with first order chains. Our focus in this article, however, is on the higher order case. It should be pointed out that not all questions related to higher order chains can be addressed through first order chains \cite{Han24a, Han25}. Consequently, the HOMC package is not superseded by any software tools that can only handle first order chains.

The MATLAB source files of the HOMC package are available publicly at \url{https://neumann.math.siu.edu/homc}.

\section{Description of Main Functions in HOMC Package}

Now, we describe the main functions in the HOMC package along with necessary and brief background information. These functions are for general purposes and are useful in many situations. We also provide examples to illustrate how these functions can be used.

\subsection{Linear Indexing}

Linear indexing \cite{Martin13} is one way of ordering $(i_1, i_2, \ldots, i_m)$ with $i_1, i_2, \ldots, i_m \in S$. Simply put, it means that we run these indices from $1$ through $n$ in the precedence from left to right. The reversed linear indexing, on the other hand, means in the precedence from right to left. The function for these tasks is 
\begin{center}
\framebox{{\tt ind= lind(s,r)},}
\end{center}
where {\tt s} is the row vector 
\begin{equation}
\label{s}
{\tt s} =[\underbrace{n \ \ \ n \ \ \ \ldots \ \ \ n}_m]
\end{equation}
and {\tt r} is optional unless the reversed linear indexing is needed. Observe that for an $(m-1)$th order chain on state space $S=\{1, 2, \ldots, n\}$, {\tt s} is exactly the size of its transition tensor $\mathcal{P}$ and can be easily obtained in MATLAB by {\tt s=size(P)}. The default value of {\tt r} is $1$, which leads to the linear indexing of $(i_1, i_2, \ldots, i_m)$ for $i_1, i_2, \ldots, i_m \in S$. The reversed linear indexing is achieved by setting {\tt r} to $-1$.
\begin{6}
Let $s=[2 \ \ \ 2 \ \ \ 2 \ \ \ 2]$. Then, {\tt ind=lind(s)} gives: 
\begin{verbatim}
     1     1     1     1
     2     1     1     1
     1     2     1     1
     2     2     1     1
     1     1     2     1
     2     1     2     1
     1     2     2     1
     2     2     2     1
     1     1     1     2
     2     1     1     2
     1     2     1     2
     2     2     1     2
     1     1     2     2
     2     1     2     2
     1     2     2     2
     2     2     2     2
\end{verbatim}
while {\tt ind=lind(s,-1)} gives: 
\begin{verbatim}
     1     1     1     1
     1     1     1     2
     1     1     2     1
     1     1     2     2
     1     2     1     1
     1     2     1     2
     1     2     2     1
     1     2     2     2
     2     1     1     1
     2     1     1     2
     2     1     2     1
     2     1     2     2
     2     2     1     1
     2     2     1     2
     2     2     2     1
     2     2     2     2
\end{verbatim}
\end{6}
Clearly, the reversed linear indexing can be obtained from linear indexing by MATLAB's {\tt fliplr}. This is what {\tt lind} does when {\tt r} is set to $-1$.

\subsection{Tensor $\boxtimes$ Product and Powers}

Given two $m$th order, $n$ dimensional tensors $\mathcal{A}=[a_{i_1i_2\ldots i_m}]$ and $\mathcal{B}=[b_{i_1i_2\ldots i_m}]$, $\mathcal{A} \boxtimes \mathcal{B}$ is defined \cite{Han24a, Han25} to be an $m$th order, $n$ dimensional tensor $\mathcal{C}=[c_{i_1i_2\ldots i_m}]$ such that 
$$c_{i_1i_2\ldots i_m}=\sum_{j=1}^n a_{i_1ji_2\ldots i_{m-1}}b_{ji_2\ldots i_m}$$
for all $1 \le i_1, i_2, \ldots, i_m \le n$. As a special case, when $\mathcal{A}$ and $\mathcal{B}$ are both $n$ by $n$ matrices, $\mathcal{A} \boxtimes \mathcal{B}$ reduces to the usual matrix multiplication $\mathcal{AB}$. Unlike its matrix counterpart, however, the $\boxtimes$ product is not associative when $m \ge 3$. In other words, for $m$th order, $n$ dimensional tensors $\mathcal{A}$, $\mathcal{B}$, and $\mathcal{C}$, $$\mathcal{A} \boxtimes (\mathcal{B} \boxtimes \mathcal{C}) \ne (\mathcal{A} \boxtimes \mathcal{B}) \boxtimes \mathcal{C}$$
 in general.
 
The function for computing $\mathcal{A} \boxtimes \mathcal{B}$ can be called by  
\begin{center}
\framebox{{\tt C=bprod(A,B)},}
\end{center}
where {\tt A}, {\tt B}, and {\tt C} are self-explanatory.

The power of an $m$th order, $n$ dimensional tensor $\mathcal{A}$ is defined \cite{Han24a, Han25} recursively according to 
$$\mathcal{A}^{k+1}=\mathcal{A}^k \boxtimes \mathcal{A}, ~k=0, 1, 2, \ldots,$$
where, in particular, $\mathcal{A}^0$ is the $m$th order, $n$ dimensional identity tensor $\mathcal{I}=[\delta_{i_1i_2\ldots i_m}]$ with 
$$\delta_{i_1i_2\ldots i_m}=\left\{\begin{array}{cl}
1, & i_1=i_2;\\
0, & {\rm otherwise}.
\end{array}\right.$$
Clearly, $\mathcal{I} \boxtimes \mathcal{A}=\mathcal{A}$ but, in general, $\mathcal{A} \boxtimes \mathcal{I} \ne \mathcal{A}$. Observe that this identity tensor is different from those in \cite{Martin13, Qi17}.

To compute the $k$th power of $\mathcal{A}$, we can use the function 
\begin{center}
\framebox{{\tt C=bpow(A,k)}.}
\end{center}
We comment that because the $\boxtimes$ product is not associative, $\mathcal{A}^6$, for example, cannot be computed as $\mathcal{A}^3 \boxtimes \mathcal{A}^3$. To form the $m$th order, $n$ dimensional identity tensor, the command is 
\begin{center}
\framebox{{\tt I=eyet(s)},}
\end{center}
where {\tt s} is as in (\ref{s}).

The $\boxtimes$ product and power play a key role in the study of higher order Markov chains \cite{Han24a, Han24b, Han25}. To begin with, denote $\mathcal{P}^k=[p^{(k)}_{i_1i_2\ldots i_m}]$, where $\mathcal{P}$ is the transition tensor and $k \ge 1$. Then, $p^{(k)}_{i1i_2\ldots i_m}$ is the $k$-step transition probability from states $(i_2, \ldots, i_m)$ to $i_1$, that is,
$$p^{(k)}_{i_1i_2\ldots i_m}=\Pr(X_{t+k}=i_1 | X_t=i_2, \ldots, X_{t-m+2}=i_m).$$ Accordingly, $\mathcal{P}^k$ is called the $k$-step transition tensor. Clearly, $\mathcal{P}^1=\mathcal{P}$.

\begin{6}
\label{reg}
Consider a second order chain on $S=\{1, 2, 3, 4\}$ with transition tensor 
$$\mathcal{P}(:,:,1)=\left[\begin{array}{cccc}
0.5 & 0 & 0 & 0\\
0.5 & 0 & 1 & 0\\
0 & 1 & 0 & 1\\
0 & 0 & 0 & 0
\end{array}\right], ~\mathcal{P}(:,:,2)=\left[\begin{array}{cccc}
0 & 0 & 0.5 & 1\\
0 & 0.5 & 0 & 0\\
0.5 & 0.5 & 0 & 0\\
0.5 & 0 & 0.5 & 0
\end{array}\right],$$
$$\mathcal{P}(:,:,3)=\left[\begin{array}{cccc}
0 & 1 & 0 & 1\\
1 & 0 & 0.5 & 0\\
0  & 0 & 0.5 & 0\\
0 & 0 & 0 & 0
\end{array}\right], ~\mathcal{P}(:,:,4)=\left[\begin{array}{cccc}
0 & 0 & 0 & 0\\
1 & 1 & 1 & 0\\
0 & 0 & 0 & 0.5\\
0 & 0 & 0 & 0.5
\end{array}\right].$$
Then, using {\tt bpow(P,10)}, we see 
$$\mathcal{P}^{10}(:,:,1)=\left[\begin{array}{cccc}
    0.2959  &  0.3750  &  0.2500  &  0.5000\\
    0.2158  &  0.3750  &  0.1250  &  0.2500\\
    0.3066  &  0.1250  &  0.3750  &  0.1250\\
    0.1816  &  0.1250  &  0.2500  &  0.1250
    \end{array}\right],$$
     $$\mathcal{P}^{10}(:,:,2)=\left[\begin{array}{cccc}
    0.3750  &  0.3350  &  0.1875  &  0.1250\\
    0.3750  &  0.2959  &  0.3750  &  0.2500\\
    0.1875  &  0.2158  &  0.3750  &  0.5000\\
    0.0625  &  0.1533  &  0.0625  &  0.1250  
    \end{array}\right],$$  
$$\mathcal{P}^{10}(:,:,3)=\left[\begin{array}{cccc}
    0.3750  &  0.1250  &  0.3066  &  0.1250\\
    0.1250  &  0.3750  &  0.2158  &  0.2500\\
    0.2500  &  0.3750  &  0.2959  &  0.5000\\
    0.2500  &  0.1250  &  0.1816  &  0.1250
    \end{array}\right],$$ 
    $$\mathcal{P}^{10}(:,:,4)=\left[\begin{array}{cccc}
    0.3750  &  0.2949  &  0.2500  &  0.3633\\
    0.1250  &  0.2168  &  0.1250  &  0.3066\\
    0.2500  &  0.3066  &  0.3750  &  0.2158\\
    0.2500  &  0.1816  &  0.2500  &  0.1143  
    \end{array}\right].$$
Since there exists $k \ge 1$, i.e., $k=10$ here, such that $\mathcal{P}^k>0$, this higher order chain is called regular \cite{Han25}. An important consequence of such regularity is that the higher order chain must have a unique limiting probability distribution \cite{Han25}.
\end{6}

\begin{6}
\label{erg}
Take a second order chain on $S=\{1, 2, 3\}$ whose transition tensor is 
$$\mathcal{P}(:,:,i_3)=\left[\begin{array}{ccc}
0 & 0.5 & 0\\
1 & 0 & 1\\
0 & 0.5 & 0
\end{array}\right]$$
for $i_3=1, 2, 3$. Then, with {\tt bpow(P,k)}, it is easy to check that $\mathcal{P}^k=\mathcal{P}$ for $k=1, 3, 5, \ldots$, and  
$$\mathcal{P}(:,:,i_3)=\left[\begin{array}{ccc}
0.5 & 0 & 0.5\\
0 & 1 & 0\\
0.5 & 0 & 0.5
\end{array}\right], ~i_3=1, 2, 3,$$
for $k=2, 4, 6, \ldots$. Clearly, this chain is not regular. Given any $i_1, i_2, i_3 \in S$, however, there exists some $k \ge 1$, which usually depends on $i_1, i_2, i_3$, such that $p^{(k)}_{i_1i_2i_3}>0$. Such a higher order chain is called ergodic \cite{Han24a}. It is shown in \cite{Han24a} that when a higher order chain is ergodic, its mean first passages times are well-defined.
\end{6}

As mentioned earlier, the HOMC package works for first order chains too. Below is one such example.

\begin{6}
\label{first}
Let $$P=\left[\begin{array}{ccc}
0.5 & 0.5 & 0\\
0.5 & 0 & 1\\
0 & 0.5 & 0
\end{array}\right]$$ 
be the transition matrix of a first order chain with three states. Then, from {\tt bpow(P,5)}, we get the $5$-step transition matrix of the chain as 
$$\left[\begin{array}{ccc}
    0.3750  &  0.4688  &  0.3125\\
    0.4688  &  0.2188  &  0.6250\\
    0.1562  &  0.3125  &  0.0625
    \end{array}\right].$$
which is the same as the result from {\tt P\verb|^|5} in MATLAB, the usual way how this matrix is formed in the first order case \cite{Kemeny60}.
\end{6}

\subsection{Diagonal Tensors}

The identity tensor introduced in the previous subsection is a special diagonal tensor. More generally, if $\mathcal{A}=[a_{i_1i_2\ldots i_m}]$ is an $m$th order, $n$ dimensional tensor, then the diagonal tensor obtained from $\mathcal{A}$ is defined as 
$\mathcal{A}_d=[a^{(d)}_{i_1i_2\ldots i_m}]$ such that 
$$a^{(d)}_{i_1i_2\ldots i_m}=\left\{\begin{array}{cl}
a_{i_1i_2\ldots i_m}, & i_1=i_2;\\
0, & {\rm otherwise}.
\end{array}\right.$$
In other words, $\mathcal{A}_d$ is derived from $\mathcal{A}$ by setting all its off-diagonal entries to zero. We comment that the diagonal tensor here is different from those in \cite{Kolda09, Qi17}.

To extract the diagonal tensor $\mathcal{A}_d$ from $\mathcal{A}$, we simply call 
\begin{center}
\framebox{{\tt D=diagt(A)}.}
\end{center}
This function is useful for dealing with problems such as ever-reaching probabilities and mean first passage times as we shall see in Section \ref{add}.

\subsection{Matricization and Tensorization}

In certain situations, it is more convenient to represent a tensor as a matrix or vice versa. These are called matricization and tensorization, respectively. In such conversions, linear indexing is usually adopted for consistency of results.

Suppose that $\mathcal{A}=[a_{i_1i_2\ldots i_m}]$ is an $m$th order, $n$ dimensional tensor. Let $N=n^{m-1}$. Then, for any $1 \le k \le n$, the mode-$k$ matricization of $\mathcal{A}$ is an $n$ by $N$ matrix $B=[b_1 ~b_2 ~\ldots ~b_N]$ in columewise form such that its columns are the mode-$k$ fibers of $\mathcal{A}$, namely, $\mathcal{A}(i_1, \ldots, i_{k-1}, 1:n, i_{k+1}, \ldots, i_m)$, and are arranged via the linear indexing order of $(i_1, \ldots, i_{k-1}, i_{k+1}, \ldots, i_m)$ \cite{Kolda09}. The converse of this process, on the other hand, produces the mode-$k$ tensorization, written as $\mathcal{A}$ for example, of an $n$ by $N$ matrix $B=[b_1 ~b_2 ~\ldots ~b_N]$.

The mode-$k$ matricization and its reverse tensorization are done by the functions 
\begin{center}
\framebox{{\tt B=t2mat(A,k)}}
\end{center}
and 
\begin{center}
\framebox{{\tt A=mat2t(B,k)},}
\end{center}
respectively. Observe that the size of $B$ must be $n$ by $n^{m-1}$ for some $n, m \ge 2$. In addition, when $\mathcal{A}$ is an $n$ by $n$ matrix, {\tt t2mat(A,1)} gives $\mathcal{A}$ and {\tt t2mat(A,2)} gives $\mathcal{A}^T$.

\begin{6}
Let $\mathcal{A}={\tt reshape(1:16,2,2,2,2)}$, i.e.,
$$\mathcal{A}(:,:,1,1)=\left[\begin{array}{cc}
1 & 3\\
2 & 4
\end{array}\right], ~\mathcal{A}(:,:,2,1)=\left[\begin{array}{cc}
5 & 7\\
6 & 8
\end{array}\right],$$
$$\mathcal{A}(:,:,1,2)=\left[\begin{array}{cc}
9 & 11\\
10 & 12
\end{array}\right], ~\mathcal{A}(:,:,2,2)=\left[\begin{array}{cc}
13 & 15\\
14 & 16
\end{array}\right].$$
Then, by {\tt B=t2mat(A,3)}, we arrive at the following mode-$3$ matricization of $\mathcal{A}$: 
$$B=\left[\begin{array}{cccccccc}
     1  &   2  &   3  &   4  &   9  &  10  &  11  &  12\\
     5  &   6  &   7  &   8  &  13  &  14  &  15  &  16
     \end{array}\right].$$
In addition, the command {\tt A=mat2t(B,3)} yields back $\mathcal{A}$.
\end{6}

Applications of matricization to the study of higher order chains will be demonstrated soon.

\subsection{Reduced First Order Chain}

It is  well known that a higher order chain can be associated with a first order chain \cite{Doob90}. Even though not all problems regarding a higher order chain can be solved by its associated first order chain \cite{Han24a, Han25}, there are circumstances where such an associated first order chain is quite useful, especially when it comes to the problem of finding the limiting probability distribution \cite{Han25}.

Given the higher order chain $X=\{X_t : t=1, 2, \ldots\}$ with an $m$th order, $n$ dimensional transition tensor $\mathcal{P}$ as laid out in the introductory section, define $T=\{i_1i_2\ldots i_{m-1} : i_1, i_2, \ldots, i_{m-1} \in S\}$ and $Y_t=[X_t ~X_{t-1} ~\ldots ~X_{t-m+2}]^T$ for $t \ge m-1$. The set $T$ consists of all $N=n^{m-1}$ multi-indices $i_1i_2\ldots i_{m-1}$ of length $m-1$, ordered according to linear indexing of $(i_1, i_2, \ldots, i_{m-1})$. By identifying $Y_t=[i_1 ~i_2 ~\ldots ~i_{m-1}]$ with $Y_t=i_1i_2\ldots i_{m-1}$, the vector-valued stochastic process $Y=\{Y_t : t=m-1, m, \ldots\}$ turns out to be a first order chain on state space $T$ \cite{Doob90}. It is called \cite{Han25} the reduced first order chain obtained from the higher order chain $X$.

The probability distribution of the chain $X$ at $t$ is defined to be 
\begin{equation}
\label{xt}
x_t=[\Pr(X_t=1) \ \ \ \Pr(X_t=2) \ \ \ \ldots \ \ \ \Pr(X_t=n)]^T.
\end{equation}
Similarly, the probability distribution of the reduced first order chain $Y$ at $t$ is defined to be an $N$ by $1$ column vector $y_t$ whose, in multi-index form, $i_1i_2\dots i_{m-1}$th entry is $\Pr(Y_t=i_1i_2\ldots i_{m-1})$. Then, it is known \cite{Li14}
$$x_{t+1}=Py_t,$$
where $P$ is the mode-$1$ matricization of $\mathcal{P}$. This is where matricization in the previous subsection can be useful since it establishes a connection here between the probability distributions $x_t $ and $y_t$.

Another usage of matricization is for the construction of the $N$ by $N$ transition matrix $Q$ of the reduced first order chain $Y$, see \cite{Han25}. We give a brief description as follows.

Let 
$$G=[\underbrace{I_{n^{m-2}} \ \ \ I_{n^{m-2}} \ \ \ \ldots \ \ \ I_{n^{m-2}}}_n],$$
where $I_{n^{m-2}}$ is the $n^{m-2}$ by $n^{m-2}$ identity matrix. Then, 
$$Q=G \ast P,$$
where $\ast$ stands for the columnwise Khatri-Rao product \cite{Smilde04}.

Accordingly, there are two functions related to the formation of $Q$, one is 
\begin{center}
\framebox{{\tt krprod(A,B)},}
\end{center} 
which computes the columnwise Khatri-Rao product of matrices $A$ and $B$, and the other is \begin{center}
\framebox{{\tt Q=rcmat(P)},}
\end{center}
which takes the transition tensor $\mathcal{P}$ as input, calls {\tt t2mat} and {\tt krprod}, and returns the matrix $Q$. In the special case of a first order chain with transition matrix $P$, the output from above is the same as $P$.

\begin{6}
Let us revisit the second order chain in Example \ref{reg}. Using {\tt Q=rcmat(P)}, we obtain 
$$Q=\left[\begin{array}{cccccccccccccccc}
    0.5    &     0    &     0    &     0   &      0    &     0   &      0    &     0    &     0    &     0    &     0   &      0    &     0    &     0    &     0    &     0\\
    0.5    &     0    &     0    &     0   &      0    &     0   &      0    &     0    &     1    &     0    &     0   &      0    &     1    &     0    &     0    &     0\\
       0    &     0    &     0    &     0   &   0.5    &     0   &      0    &     0    &     0    &     0    &     0   &      0    &     0    &     0    &     0    &     0\\
       0     &    0    &     0    &     0   &   0.5    &     0   &      0    &     0    &     0    &     0    &     0   &      0    &     0    &     0    &     0    &     0\\
       0     &    0    &     0    &     0   &      0    &     0   &      0    &     0    &     0    &     1    &     0   &      0    &     0    &     0    &     0    &     0\\
       0     &    0    &     0    &     0   &      0    &  0.5   &      0    &     0    &     0    &     0    &     0   &      0    &     0    &     1    &     0    &     0\\
       0     &    1    &     0    &     0   &      0    &  0.5   &      0    &     0    &     0    &     0    &     0   &      0    &     0    &     0    &     0    &     0\\
       0     &    0    &     0    &     0   &      0    &     0   &      0    &     0    &     0    &     0    &     0   &      0    &     0    &     0    &     0    &     0\\
       0     &    0    &     0    &     0   &      0    &     0   &    0.5   &     0    &     0    &     0    &     0   &      0    &     0    &     0    &     0    &     0\\
       0     &    0    &     1    &     0   &      0    &     0   &       0   &     0    &     0    &     0    &   0.5  &      0    &     0    &     0    &      1   &     0\\
       0     &    0    &     0    &     0   &      0    &     0   &       0   &     0    &     0    &     0    &   0.5  &      0    &     0    &     0    &      0   &     0\\
       0     &    0    &     0    &     0   &      0    &     0   &    0.5   &     0    &     0    &     0    &      0  &      0    &     0    &     0    &      0   &     0\\
       0     &    0    &     0    &     0   &      0    &     0   &       0   &     1    &     0    &     0    &      0  &      1    &     0    &     0    &      0   &     0\\
       0     &    0    &     0    &     0   &      0    &     0   &       0   &     0    &     0    &     0    &      0  &      0    &     0    &     0    &      0   &     0\\
       0     &    0    &     0    &     1   &      0    &     0   &       0   &     0    &     0    &     0    &      0  &      0    &     0    &     0    &      0   &  0.5\\
       0     &    0    &     0    &     0   &      0    &     0   &       0   &     0    &     0    &     0    &      0  &      0    &     0    &     0    &      0   &  0.5
        \end{array}\right].$$
It is shown in \cite{Han25} that when a higher order chain is regular, its limiting probability distribution $\pi=\lim_{t \rightarrow \infty}x_t$, where $x_t$ is given in (\ref{xt}), exists and is unique. In addition, $\pi$ can be determined by 
\begin{equation}
\label{pi}
\pi=P^{(0)}y,
\end{equation}
where $P^{(0)}$ is the mode-$1$ matricization of the $m$th order, $n$ dimensional identity tensor and $y$ is any right eigenvector corresponding to the dominant eigenvalue $\lambda=1$ of $Q$, normalized such that $y \ge 0$ and $\|y\|_1=1$. With MATLAB's {\tt eig} function, we find that $\lambda=1$ has multiplicity $2$ and its eigenspace is spanned by 
$$y^{(1)}=[     0         \ \ \ 0    \ \ \ 0.1429    \ \ \ 0.1429    \ \ \ 0.2857         \ \ \ 0         \ \ \ 0         \ \ \ 0         \ \ \ 0    \ \ \ 0.2857         \ \ \ 0         \ \ \ 0         \ \ \ 0         \ \ \ 0    \ \ \ 0.1429         \ \ \ 0]^T$$
and
$$y^{(2)}=[     0    \ \ \ 0.2857         \ \ \ 0         \ \ \ 0         \ \ \ 0         \ \ \ 0    \ \ \ 0.2857         \ \ \ 0    \ \ \ 0.1429         \ \ \ 0         \ \ \ 0    \ \ \ 0.1429    \ \ \ 0.1429         \ \ \ 0         \ \ \ 0         \ \ \ 0]^T.$$
Because of (\ref{pi}), both $y^{(1)}$ and $y^{(2)}$ lead to 
$$\pi=[   0.2857    \ \ \ 0.2857    \ \ \ 0.2857    \ \ \ 0.1429]^T.$$
\end{6}

The examples being discussed thus far have demonstrated that the HOMC package, together with a few MATLAB's built-in functions, can be a handy tool in numerical experimentation and quick prototyping on higher order chains.

\section{Additional Functions in HOMC Package}
\label{add}

Next, let us explore some more advanced and specialized functions in the HOMC package. We continue to use the settings in the introductory section for the higher order chain $X$ on state space $S=\{1, 2, \ldots, n\}$ with an $m$th order, $n$ dimensional transition tensor $\mathcal{P}=[p_{i_1i_2\ldots i_m}]$.

\subsection{Ever-Reaching Probabilities}

Let $\eta_{i_1i_2\ldots i_m}=\min\{j \ge 1 : X_{m+j-1}=i_1 | X_{m-1}=i_2, \ldots, X_1=i_m\}$ be the random variable of the first passage time from states $(i_2, \ldots, i_m)$ to $i_1$. The probability that such a first passage occurs at $j=k$ is denoted by $f^{[k]}_{i_1i_2\ldots i_m}=\Pr(\eta_{i_1i_2\ldots i_m}=k)$. It is known \cite{Han24a} that $f^{[1]}_{i_1i_2\ldots i_m}=p_{i_1i_2\ldots i_m}$ and 
$$f^{[k+1]}_{i_1i_2\ldots i_m}=\sum_{j \in S, j \ne i_1}f^{[k]}_{i_1ji_2\ldots i_{m-1}}p_{ji_2\ldots i_m}, ~k=1, 2, \ldots.$$ 
Denote $\mathcal{F}^{[k]}=[f^{[k]}_{i_1i_2\ldots i_m}]$ as a tensor. It follows that $\mathcal{F}^{[1]}=\mathcal{P}$ and, with the $\boxtimes$ product notation, 
$$\mathcal{F}^{[k+1]}=(\mathcal{F}^{[k]}-\mathcal{F}^{[k]}_d)\boxtimes \mathcal{P}, ~k=1, 2, \ldots.$$
The ever-reaching probability tensor $\mathcal{F}=[f_{i_1i_2\ldots i_m}]$ is defined to be \cite{Han24a}
\begin{equation}
\label{erp}
\mathcal{F}=\sum_{k=1}^\infty \mathcal{F}^{[k]},
\end{equation}
where the summation is in the entrywise sense, exactly in the way how MATLAB handles it.

As an application, the diagonal entries of $\mathcal{F}$ can be used to classify the states of the chain \cite{Han24b}. Specifically, a state $i$ is called recurrent if $f_{iii_3\ldots i_m}=1$ for all $i_3, \ldots, i_m \in S$; otherwise, it is called transient. In addition, a state $i$ is called fully transient if $f_{iii_3\ldots i_m}<1$ for all $i_3, \ldots, i_m \in S$. Incidentally, if the chain is ergodic, then $\mathcal{F}=\mathcal{E}$, the $m$th order, $n$ dimensional tensor of all ones \cite{Han24a}. In this case, all the states are known to be recurrent \cite{Han24b}.

Based on the function {\tt bprod}, we have implemented (\ref{erp}) in the HOMC package as a function 
\begin{center}
\framebox{{\tt F=erp(P,tol)},}
\end{center}
where {\tt tol} is the tolerance and is optional. The default value of {\tt tol} is $10^{-6}$. If for some $k \ge 1$, the largest entry in absolute value of $\mathcal{F}^{[k]}$ is less than {\tt tol}, then {\tt erp} returns the sum of the first $k$ terms in (\ref{erp}) as an approximation to $\mathcal{F}$.

\begin{6}
Consider a second order chain on $S=\{1, 2, 3\}$ whose transition tensor is given by
$$\mathcal{P}(:,:,1)=\left[\begin{array}{ccc}
1/2 & 1/3 & 1/2\\
1/2 & 1/3 & 0\\
0 & 1/3 & 1/2
\end{array}\right], ~\mathcal{P}(:,:,2)=\left[\begin{array}{ccc}
1 & 0 & 1/2\\
0 & 1 & 1/2\\
0 & 0 & 0
\end{array}\right],$$
$$\mathcal{P}(:,:,3)=\left[\begin{array}{ccc}
0 & 0 & 1/2\\
1 & 0 & 1/2\\
0 & 1 & 0
\end{array}\right].$$
Using {\tt erp(P,tol)} with {\tt tol}$=10^{-8}$ and MATLAB's {\tt format} {\tt long}, we obtain from the first $67$ terms in (\ref{erp}) that 
$$\mathcal{F}(:,:,1)=\left[\begin{array}{ccc}
   0.833333333313931 &  0.666666666627862 &  0.999999999941792\\
   1.000000000000000 &  1.000000000000000 &  1.000000000000000\\
   0.499999975649230 &  0.499999986536366 &  0.749999988833857
   \end{array}\right],$$ 
    $$\mathcal{F}(:,:,2)=\left[\begin{array}{ccc}
   1.000000000000000 &                  0 &  0.999999999941792\\
   1.000000000000000 &  1.000000000000000 &  1.000000000000000\\
   0.499999968638406 &                  0 &  0.749999988833857
   \end{array}\right],$$ 
    $$\mathcal{F}(:,:,3)=\left[\begin{array}{ccc}
   0.666666666627862 &  0.999999999883585 &  0.999999999941792\\
   1.000000000000000 &  1.000000000000000 &  1.000000000000000\\
   0.499999982660054 &  1.000000000000000 &  0.749999988833857
   \end{array}\right].$$
Besides, we see from the diagonal entries of $\mathcal{F}$ that state $1$ is transient but not fully transient, state $2$ is recurrent, and state $3$ is fully transient. 
\end{6}

\subsection{Mean First Passage Times}

Mean first passage times are important characteristics of a chain since they provide useful information about the short-term dynamics of the stochastic process. Recall that the first passage time random variable from states $(i_2, \ldots, i_m)$ to $i_1$ is denoted by $\eta_{i_1i_2\ldots i_m}$. The mean first passage time from states $(i_2, \ldots, i_m)$ to $i_1$ is defined as \cite{Han22}
$$\mu_{i_1i_2\ldots i_m}={\rm E}(\eta_{i_1i_2\ldots i_m})=\sum_{k=1}^\infty kf^{[k]}_{i_1i_2\ldots i_m}.$$
The $m$th order, $n$ dimensional tensor $\mu=[\mu_{i_1i_2\ldots i_m}]$ is called the mean first passage time tensor. It is shown in \cite{Han24a} that when the chain is ergodic, $\mu$ is uniquely determined by 
\begin{equation}
\label{mfpcomp}
\mu_{i_1i_2\ldots i_m}=1+\sum_{j \in S, j \ne i_1}\mu_{i_1ji_2\ldots i_{m-1}}p_{ji_2\ldots i_m}, ~i_1, i_2, \ldots, i_m \in S.
\end{equation}
With the $\boxtimes$ notation, the above can be written as a tensor equation 
\begin{equation}
\label{mfp}
\mu=\mathcal{E}+(\mu-\mu_d)\boxtimes \mathcal{P},
\end{equation}
where $\mathcal{E}$ is the $m$th order, $n$ dimensional tensor of all ones. We comment that $\mathcal{E}$ can be easily formed in MATLAB by {\tt ones(s)} with  {\tt s=size(P)}.

The mean first passage time tensor $\mu$ can be computed by either a direct method or an iterative method \cite{Han24a}. We now give a brief description of each of these approaches as follows.

To compute $\mu$ directly, we treat (\ref{mfpcomp}) as a linear system. It is shown in \cite{Han24a} that when the chain is ergodic, this system is nonsingular. Moreover, by arranging the unknown $\mu_{i_1i_2\ldots i_m}$ in the reversed linear indexing order, the coefficient matrix of the system in (\ref{mfpcomp}) has an $n$ by $n$ block diagonal structure, in which the $k$th diagonal block corresponds to $i_1=k$ in (\ref{mfpcomp}), $k=1, 2, \ldots, n$. This splits (\ref{mfpcomp}) naturally into $n$ smaller subsystems. We have implemented this approach by the function 
\begin{center}
\framebox{{\tt mu=mfptd(P)}.}
\end{center}
It takes the transition tensor $\mathcal{P}$ as input, solves each smaller subsystem from (\ref{mfpcomp}), and piece together these solutions to form $\mu$. We comment that it is not advisable to solve (\ref{mfpcomp}) directly as an entire $n^m$ by $n^m$ system without exploiting its underlying block diagonal structure, especially when the chain is large.

To compute $\mu$ iteratively, we rewrite (\ref{mfp}) as 
$$\mu^{(k+1)}=\mathcal{E}+(\mu^{(k)}-\mu^{(k)}_d)\boxtimes \mathcal{P}, ~k=0, 1, 2, \ldots .$$
It is shown in \cite{Han24a} that this iterative algorithm is guaranteed to converge when the chain is ergodic. A good choice of $\mu^{(0)}$ is $\mathcal{E}$ since $\mu_{i_1i_2\ldots i_m} \ge 1$ for any $i_1, i_2, \ldots, i_m \in S$ \cite{Han24a}. The algorithm terminates when the largest entrywise difference in absolute value between two successive iterations of $\mu$ is within a prescribed tolerance. We have also implemented this approach by the function 
\begin{center}
\framebox{{\tt mu=mfpti(P,mu0,tol)},}
\end{center}
where {\tt P} is the transition tensor, {\tt mu0} is the initial approximation to $\mu$, and {\tt tol} is the tolerance. The default {\tt mu0} is $\mathcal{E}$ while the default {\tt tol} is $10^{-6}$. Both {\tt mu0} and {\tt tol} are optional, but both must be specified if any of them needs to be set differently from its default value. 

\begin{6}
Now we revisit the ergodic second order chain in Example \ref{erg}. First, we use {\tt mu=mfptd(P)} to compute $\mu$. The result is 
$$\mu(:,:,i_3)=\left[\begin{array}{ccc}
     4   &  3  &   4\\
     1   &  2  &   1\\
     4   &  3  &   4
     \end{array}\right], ~i_3=1, 2, 3.$$
To check this answer, according to (\ref{mfp}), we compute
$$\mu-\mathcal{E}-(\mu-\mu_d)\boxtimes \mathcal{P}$$
by {\tt mu-ones(s)-bprod(mu-diagt(mu),P)} with {\tt s=size(P)}, which yields a zero tensor as we expect. Next, we use {\tt mu=mfpti(P)} to compute $\mu$. The algorithm converges in $40$ iterations and the output in {\tt format} {\tt long} is 
$$\mu(:,:,i_3)=\left[\begin{array}{ccc}
   3.999997138977051 &  2.999998092651367 &  3.999997138977051\\
   1.000000000000000 &  2.000000000000000 &  1.000000000000000\\
   3.999997138977051 &  2.999998092651367 &  3.999997138977051
   \end{array}\right],$$
   for $i_3=1, 2, 3$.
\end{6}

Finally, let us give one more example to illustrate the applicability of the HOMC package to first order chains. It is known \cite{Kemeny60} that for an ergodic first order chain with $n$ by $n$ transition matrix $P$, its mean first passage matrix $M$ satisfies 
$$M=E+(M-M_d)P,$$
where $E$ is the $n$ by $n$ matrix of all ones and $M_d$ is the diagonal matrix obtained from $M$ by setting all its off-diagonal entries to zero. As pointed out in \cite{Han24a}, however, the above does not yield (\ref{mfp}) although the reverse is obviously true. As a matter of fact, not all problems involving higher order chains can be solved by resorting to first order chains. Thus a package such as HOMC cannot be replaced by one that deals only with first order chains.

\begin{6}
Let us revisit the first order chain in Example \ref{first}. By {\tt M=mfptd(P)}, we obtain 
$$M=\left[\begin{array}{ccc}
    2.5000  &  3.0000  &  4.0000\\
    2.0000  &  2.5000  &  1.0000\\
    6.0000  &  4.0000  &  5.0000
\end{array}\right].$$
On the other hand, using {\tt M=mfpti(P)} with the default {\tt tol} and {\tt format} {\tt long}, we find 
$$M=\left[\begin{array}{ccc}
   2.499999999825377 &  2.999999999767169 &  3.999999999650754\\
   2.000000000000000 &  2.500000000000000 &  1.000000000000000\\
   5.999995824987868 &  3.999997419700599 &  4.999996810574538
    \end{array}\right]$$
in $66$ iterations.
\end{6}

\section{Conclusions}

In this article, we have presented a MATLAB package for higher order Markov chains. Coupled with a few built-in functions in MATLAB such as {\tt size}, {\tt ones}, and {\tt eig}, this package can be used to conveniently compute all the important quantities relevant to higher order chains in our recent works \cite{Han22, Han24a, Han24b, Han25}, including the $k$-step transition tensor, limiting probability distribution, ever-reaching probability tensor, and mean first passage time tensor. It can also be used to check the ergodicity or regularity of a chain, to construct the transition matrix of the reduced first order chain, and to classify the states of the chain \cite{Han24b, Han25}. These functionalities are useful to the research community in both theoretical investigation on and practical application of higher order Markov chains.

\end{document}